\documentclass[twocolumn,english,prl]{revtex4}
\usepackage[T1]{fontenc}
\usepackage[latin9]{inputenc}
\usepackage{graphicx}

\makeatletter

\providecommand{\tabularnewline}{\\}

\@ifundefined{textcolor}{}
{%
 \definecolor{BLACK}{gray}{0}
 \definecolor{WHITE}{gray}{1}
 \definecolor{RED}{rgb}{1,0,0}
 \definecolor{GREEN}{rgb}{0,1,0}
 \definecolor{BLUE}{rgb}{0,0,1}
 \definecolor{CYAN}{cmyk}{1,0,0,0}
 \definecolor{MAGENTA}{cmyk}{0,1,0,0}
 \definecolor{YELLOW}{cmyk}{0,0,1,0}
}

\makeatother

\usepackage{babel}
\begin{document}
\title{Ground State Properties of the Diluted Sherrington-Kirkpatrick Spin
Glass}
\author{Stefan Boettcher}
\email{sboettc@emory.edu}

\affiliation{Physics Department, Emory University, Atlanta, Georgia 30322, USA}
\begin{abstract}
We present a numerical study of ground states of the dilute versions
of the Sherrington-Kirkpatrick (SK) mean-field spin glass. In contrast
to so-called ``sparse'' mean-field spin glasses that have been studied
widely on random networks of finite (average or regular) degree, the
networks studied here are randomly bond-diluted to an overall density
$p$, such that the average degree diverges as $\sim pN$ with the
system size $N$. Ground-state energies are obtained with high accuracy
for random instances over a wide range of fixed $p$. Since this is
a NP-hard combinatorial problem, we employ the Extremal Optimization
heuristic to that end. We find that the exponent describing the finite-size
corrections, $\omega$, varies continuously with $p$, a somewhat
surprising result, as one would not expect that gradual bond-dilution
would change the $T=0$ universality class of a statistical model.
For $p\to1$, the familiar result of $\omega(p=1)\approx\frac{2}{3}$
for SK is obtained. 
\end{abstract}
\maketitle
The Sherrington-Kirkpatrick model (SK) \citep{Sherrington75} was
devised as the mean-field limit of finite-dimensional Ising spin glasses,
first introduced by Edwards and Anderson (EA) \citep{Edwards75},
to describe the unusual phenomenology \citep{F+H} of disorder in
the interaction between classical dipolar magnets in certain materials.
Despite the dramatic simplification that such a limit entails, i.e.,
replacing the lattice with a dense network of bonds between all mutual
pairs of spins, SK proved so intricate that it took several years
and a herculean effort by Parisi to reveal its full structure, referred
to as replica symmetry breaking (RSB) \citep{Parisi79,Parisi80,MPV}.
RSB was verified rigorously only thirty years later \citep{Talagrand03,Panchenko12}.
Over the years, the importance of these Ising spin glass models has
significantly increased as a most concise conceptualization of systems
with disorder and frustration, and the complex structure and dynamics
that emerges \citep{MPV,Stein13}. Far beyond its origins in materials
science, SK has inspired notions of learning in neural networks and
artificial intelligence \citep{Hopfield82}, actual neurons \citep{Schneidman06},
facilitated optimization of hard combinatorial problems in operations
research and engineering \citep{MPV,MPZ,Mezard06,Boettcher19}, elucidated
the nature of energy landscapes \citep{Wales03}, made connections
to biological evolution \citep{Kauffman1989}, social dynamics \citep{Axelrod1997},
etc. Ironically, in most of these applications, the unstructured mean-field
version of a glass, such as SK, is far more realistic than the lattice
geometry of EA. Moreover, many of these problems, like optimization
and learning, concern the low-temperature limit, instead of the physically
pertinent phase transition at some finite critical temperature $T_{c}$:
As long as $T_{c}>0$, a glassy phase exists at $T\to0$. In fact,
in mean-field there is an entire critical line extending from $T_{c}$
to $T=0$ \citep{Nishimori01}. Notably, $T=0$ is its own fixed point
\citep{Oppermann07} in the renormalization group sense \citep{Pathria},
with its own set of scaling relations, yet to be completed \citep{Bouchaud03,Boettcher05d,Parisi08,BoFa11,LevySK},
connecting domain-wall excitations, ground-state energy fluctuations,
and finite-size corrections. 

Extending RSB to glassy systems on sparse networks, i.e., random graphs
\citep{Bollobas} of finite average or fixed degree (``Bethe lattices'',
BL), constituted another major breakthrough \citep{mezard:01}. More
recently, the one-dimensional long-range model \citep{Kotliar83}
has gained popularity \citep{Katzgraber03,Boettcher07b,Katzgraber2009,Aspelmeier16}
for the ability to interpolate between SK and the EA (but on a \emph{1d}-ring
geometry) based on the range of interactions. That model has effective
upper and lower dimensions, but all results obtained are numerical.

\begin{figure}
\vspace{-0.1cm}

\hfill{}\includegraphics[viewport=0bp 140bp 600bp 580bp,clip,width=1\columnwidth]{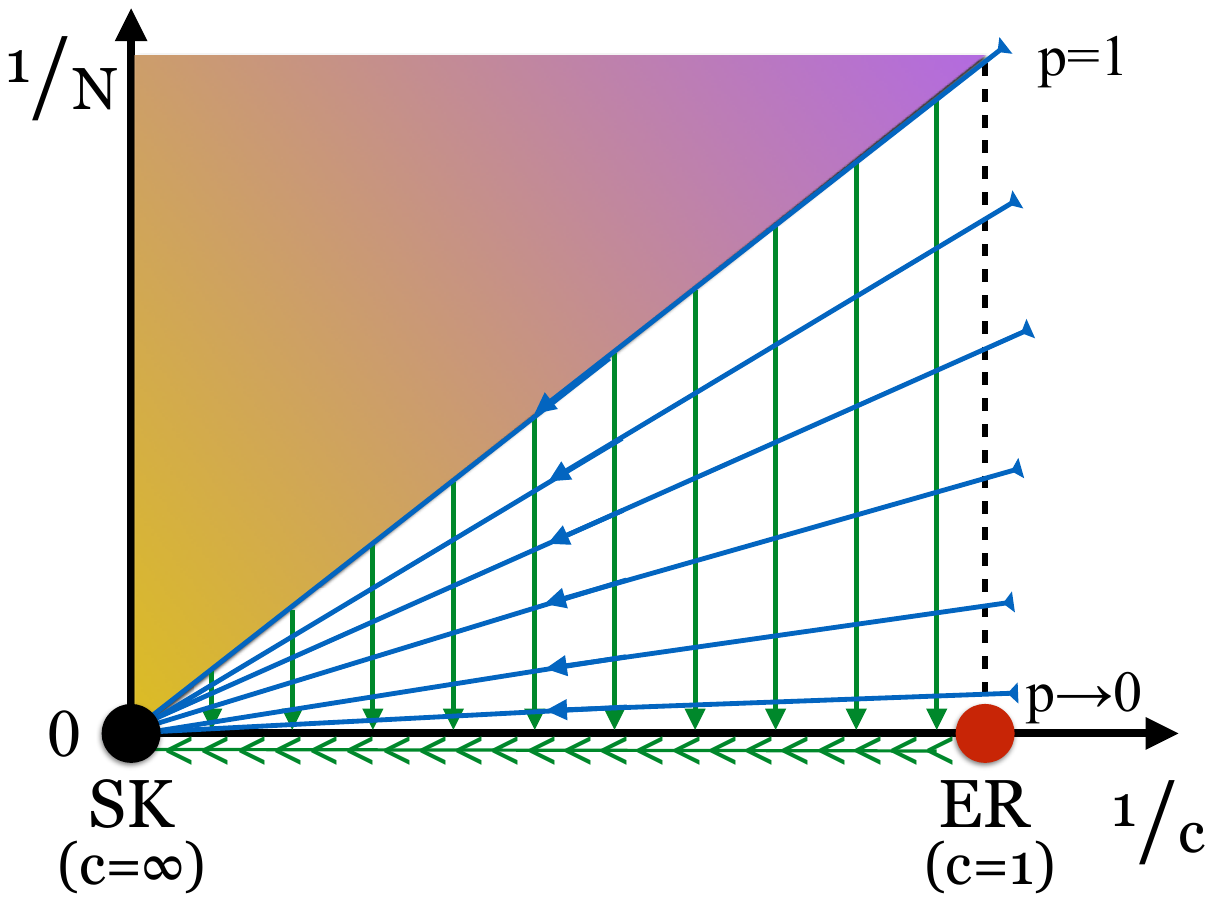}\hfill{}

\vspace{-0.2cm}

\caption{\label{fig:diluteSKphase}Depiction of alternative ways to approach
the thermodynamic limit $N\to\infty$ (or, $1/N\to0$) for mean-field
spin-glass models of (average or fixed) spin-degree $c$. Previous
work had been focused on constant $c$ while $1/N\to0$ (green down-arrows
), referred to as ``Bethe lattices'' due to their locally tree-like
structure \citep{mezard:01}. In Ref.\ \citep{Boettcher03b,Boettcher03a},
it was shown that the thermodynamic limit of their ground-state energy
densities $\left\langle e_{0}\right\rangle _{N=\infty}^{{\rm Bethe}}$
can be connected (horizontal green arrows) to that of SK (black dot)
via $\left\langle e_{0}\right\rangle _{N=\infty}^{{\rm Bethe}}\sim c^{\frac{1}{2}}e_{{\rm Parisi}}$,
at least for $c\gg1$, i.e., above the Erd\"os-R\'enyi percolation transition
for sparse random graphs \citep{Bollobas} (red dot). This study explores
a diluted SK system, in which system size $N$ and connectivity $c$
both evolve such that $p\sim c/N$ remains constant (blue rays). }
\end{figure}
It is thus surprising that after so many years of studying mean-field
spin glasses in the thermodynamic limit on fully connected (SK) or
on sparse networks (BL), there has been no consideration given to
dense but \emph{dilute} systems. (Ref.\ \citep{Fu86}, concerning
optimal graph bipartitioning, a problem closely related to spin glasses
\citep{Zdeborova10}, might pose a rare exception.) For BL, the average
or fixed number $c$ of other spins that any one spin is randomly
bonded with, i.e., its ``degree'', is held constant for all network
sizes $N\to\infty$. In contrast, it is the average bond density,
\begin{equation}
p=\frac{c}{N-1},\label{eq:diluteconnect}
\end{equation}
that is held constant in a dilute system. Clearly, in SK each spin
has a bond to every one of the other spins, i.e., $c_{{\rm SK}}=N-1$
and $p=1$, while at some general $0<p\leq1$, the degree for each
spin diverges as $c\sim pN$ in the thermodynamic limit $N\to\infty$.
Thus, dilute SK presents a true alternative to BL, for which $p\sim1/N\to0$,
likely resulting in an alternative RSB analysis. These connections
are illustrated in Fig.\ \ref{fig:diluteSKphase}. In this Letter,
we provide some tantalizing numerical evidence that such an analysis
might be quite distinct and potentially more fruitful in revealing,
for instance, the nature of finite-size corrections (FSC) that occur
when $N\to\infty$, which have remained beyond the scope of RSB.

Understanding the nature of FSC for $N\to\infty$ is an essential
ingredient in the proper interpretation of numerical data obtained
from thermodynamic systems \citep{barber:83}. To reach the thermodynamic
limit with data derived from, inevitably, finite-size simulations
usually requires a certain degree of extrapolation \citep{Palassini00,Campbell04,Boettcher03b,Boettcher10a,Boettcher10b,EOSK,Aspelmeier07,Boettcher03a}.
Here, we will specifically focus on FSC to the ensemble average of
the ground state energy density, assuming the form
\begin{equation}
\left\langle e_{0}\right\rangle _{N}\sim\left\langle e_{0}\right\rangle _{\infty}+\frac{A}{N^{\omega}},\qquad(N\to\infty),\label{eq:FSC}
\end{equation}
defining the energy density in the thermodynamic limit, $\left\langle e_{0}\right\rangle _{\infty}$.
In many disordered systems, such as for spin glasses in the low-temperature
limit exhibiting RSB, those FSC are dogged by (unknown) sub-extensive
transients \citep{Bouchaud03,BoFa11}, i.e., transients that diminish
slower than the bulk, $\omega<1$, which at times obscure the physical
interpretation to a point of arbitrariness \citep{LevySK}. Even in
mean-field, exact results for scaling properties of the glassy phase
at short of the thermodynamic limit are few \citep{Parisi08,Parisi09,Rizzo09b,parisi:93,parisi:93b}.
Finding an accessible problem as a model to make conceptual inroads
on determining FSC would thus constitute a major advance for RSB.

Numerical simulation, in fact, have provided numerous insights into
the nature of FSC in Ising spin-glass models. It was found that ground
state energies (and entropies) for mean-field systems of $N$ spins
have FSC decaying to excellent approximation with $N^{-\frac{2}{3}}$.
This was observed first for BL with bimodal bonds \citep{Boettcher03a,Boettcher03b}
and subsequently \citep{EOSK,Aspelmeier07,Kim07,Boettcher10b} also
for SK. (BL with Gaussian bonds exhibit FSC with $\omega\approx0.8$
\citep{Boettcher10a,Parisi19}.) For finite-dimensional Ising spin
glasses (EA), FSC-collapse of domain-wall excitations at $T\to0$
allowed an accurate determination of the stiffness exponent $\theta$
in dimensions $d=3,\ldots,7$ \citep{Boettcher04c}. This exponent
is fundamental to many aspects of the glassy state \citep{F+H}, for
instance, $\theta(d_{l})=0$ defines the lower critical dimension,
which appears close to $d_{l}=2.5$ \citep{Franz94,Boettcher05d,Maiorano18},
while its determination for $d\geq6$ allowed a direct check on mean-field
predictions \citep{Parisi08}. In particular, FSC were shown to decay
consistently as in Eq.\ (\ref{eq:FSC}), applied to hyper-cubic lattices
of size $N=L^{d}$ with $\omega=1-\theta/d$ \citep{BoFa11}, suggesting
the importance of domain-wall excitations for FSC \citep{Bouchaud03}.
Recently, we have proposed to use FSC analysis to assess the quality
and scalability of optimization heuristics for hard combinatorial
problems \citep{Boettcher19}.

\begin{figure*}
\hfill{}\includegraphics[viewport=0bp 0bp 700bp 618bp,clip,width=1\textwidth]{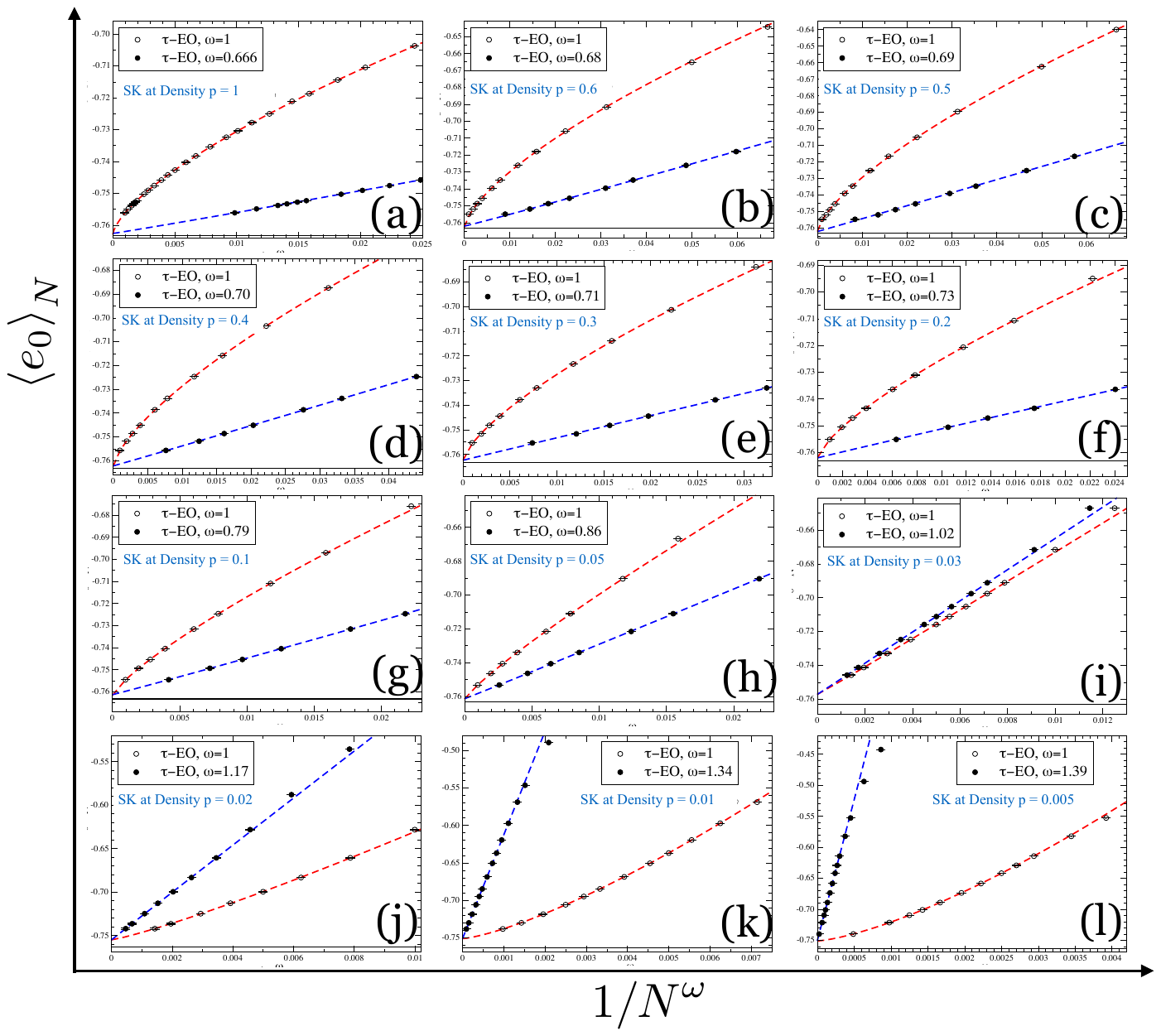}\hfill{}

\caption{\label{fig:diluteSKextra}Extrapolation for the rescaled ground-state
energy densities, $\left\langle e_{0}\right\rangle _{N}$, of the
diluted SK model of bond-density $p$ at different sizes $N$, where
each data point is plotted once for $1/N$ (i.e., $\omega=1$, open
symbols) and a second time for $1/N^{\omega}$ with a value of $\omega$
chosen such that the extrapolation to the thermodynamic limit at the
intercept $1/N^{\omega}\to0$ is asymptotically linear (closed symbols).
Each panel (a-l) depicts a different density $p$, where data is fitted
to the asymptotic form in Eq.\ (\ref{eq:FSC}) (drawn as either red
or blue-dashed lines, resp.)Each fit obtains the exponent $\omega$
and the thermodynamic ground-state energy density $\left\langle e_{0}\right\rangle _{N=\infty}$,
which should approach the Parisi energy density, $e_{{\rm Parisi}}=\left\langle e_{0}\right\rangle _{N\to\infty}$,
for all $p$ \citep{Carmona06} (horizontal line), listed in Tab.\ \ref{datatable}. }
\end{figure*}
\begin{table}
\caption{\label{datatable} List of the fitted values for the average ground
state energies $\left\langle e_{0}\right\rangle _{N=\infty}$, the
correction amplitude $A$, and the FSC exponent $\omega$ of the SK
model at various bond-densities $p$, obtained by fitting the numerical
data displayed in Fig.\ \ref{fig:diluteSKextra} to the asymptotic
form in Eq.\ (\ref{eq:FSC}). That fit was conducted over the specified
range of system sizes $N$.}
\begin{tabular}{r@{\extracolsep{0pt}.}l|r@{\extracolsep{0pt}.}lr@{\extracolsep{0pt}.}lr@{\extracolsep{0pt}.}lr@{\extracolsep{0pt}.}l}
\hline 
\multicolumn{2}{c|}{$p$} & \multicolumn{2}{c}{$\left\langle e_{0}\right\rangle _{\infty}$ } & \multicolumn{2}{c}{$\omega$} & \multicolumn{2}{c}{$A$} & \multicolumn{2}{c}{$N$-range}\tabularnewline
\hline 
\hline 
0&005 & -0&751(1)  & 1&39(1) & \multicolumn{2}{c}{448(5)} & \multicolumn{2}{c}{340-2047}\tabularnewline
0&01 & -0&752(1)  & 1&32(1) & \multicolumn{2}{c}{125(5)} & \multicolumn{2}{c}{165-2047}\tabularnewline
0&02 & -0&755(1)  & 1&16(1) & \multicolumn{2}{c}{26(3)} & \multicolumn{2}{c}{255-1023}\tabularnewline
0&03 & -0&757(1)  & 1&02(1)  & \multicolumn{2}{c}{9(1)} & \multicolumn{2}{c}{180-512}\tabularnewline
0&05 & -0&761(1)  & 0&86(1) & 3&3(5) & \multicolumn{2}{c}{165-512}\tabularnewline
0&1 & -0&762(1)  & 0&79(1) & 1&7(1) & \multicolumn{2}{c}{63-1023}\tabularnewline
0&2 & -0&762(1)  & 0&73(1) & 1&04(7) & \multicolumn{2}{c}{63-1023}\tabularnewline
0&3 & -0&762(1) & 0&71(1) & 0&91(5) & \multicolumn{2}{c}{63-1023}\tabularnewline
0&4 & -0&762(1) & 0&70(1) & 0&86(5) & \multicolumn{2}{c}{63-1023}\tabularnewline
0&5 & -0&762(1) & 0&69(1) & 0&80(4) & \multicolumn{2}{c}{45-1023}\tabularnewline
0&6 & -0&762(1) & 0&68(1) & 0&75(3) & \multicolumn{2}{c}{45-1023}\tabularnewline
1&0 & -0&763 23(5) & 0&666(3) & 0&71(1) & \multicolumn{2}{c}{80-2047}\tabularnewline
\hline 
\end{tabular}
\end{table}
\begin{figure}
\hfill{}\includegraphics[viewport=0bp 200bp 540bp 618bp,clip,width=1\columnwidth]{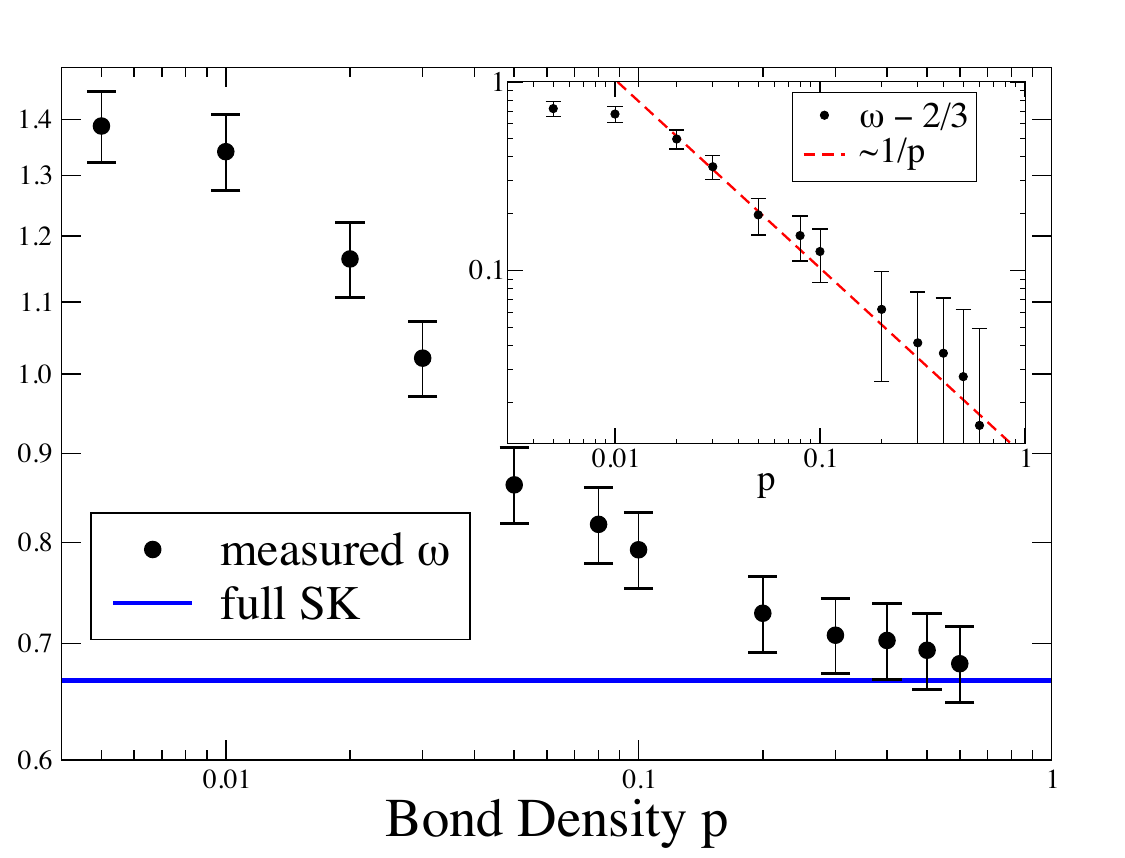}\hfill{}

\caption{\label{fig:omegaSKplot}Plot of the fitted values for the exponent
$\omega$ controlling the FSC in the extrapolation of the ground state
energies shown in Fig.\ \ref{fig:diluteSKextra} for the bond-diluted
SK model as a function of bond-density $p$. The date for $\omega$
can be found in Tab.\ \ref{datatable}. Inset: Except for the smallest
values of $p$, the exponent subtracted by its value for SK ($\omega_{{\rm SK}}\approx\frac{2}{3}$
at $p=1$), i.e., $\omega-\frac{2}{3}$, appears to approach the SK-value
roughly hyperbolically, $\sim1/p$.}
\end{figure}
In this study, we generate $N\times N$ symmetric bond matrices
with entries from a dilute bond-distribution
\begin{equation}
P\left(J\right)=p\delta\left(J^{2}-\frac{1}{pN}\right)+\left(1-p\right)\delta\left(J\right),\label{eq:PofJ}
\end{equation}
 such as to minimize the SK-Hamiltonian \citep{Sherrington75},
\begin{equation}
H_{J}=-\sum_{i>j}J_{ij}\sigma_{i}\sigma_{j},\label{eq:Hamiltonian}
\end{equation}
over the set of $N$ Ising spin variables, $\sigma_{i}=\pm1$. We
thus approximate the ground state energy density, $e_{0}(N,p)=\frac{1}{N}\min_{\vec{\sigma}}H_{J}$,
for each instance $J$. Like for SK at $p=1$, $P\left(J\right)$
is symmetric with variance $\left\langle J^{2}\right\rangle =1/N$
but higher moments for the dilute SK diverge for $p\to0$, i.e., $\left\langle J^{2n}\right\rangle =p/\left(pN\right)^{n}$
for $n=2,3,\ldots$ \footnote{As pointed out by a referee.}. For
each bond-density $p$ ($0<p\leq1$), we sample ensemble averages
$\left\langle e_{0}\right\rangle _{N}$ of the ground state energies
over a range of sizes $N$, where $P\left(J\right)$ in Eq.\ (\ref{eq:PofJ})
ensures that the average thermodynamic ground state energy is universal
\citep{Carmona06,Panchenko12}, $\left\langle e_{0}\right\rangle _{\infty}=e_{{\rm Parisi}}=-0.7631667265\ldots$
, first approximated by Parisi \citep{Parisi79}. Here, we report
on the results for a range of values $p<1$ and find surprisingly
non-trivial behavior in the continuous dependence of $\omega(p)$.
As the topology of the diagram in Fig.\ \ref{fig:diluteSKphase}
suggests, RSB should remain in effect for all $p$, possibly even
in the limit $p\to0$, where a solution should become trivial. (Even
for the smallest constant $p$, there is a neighborhood of the thermodynamic
limit, for sizes $\frac{1}{p}\ll N<\infty$ , where the dilute system
is dense enough to be above the percolation transition for sparse
random graphs at $c=1$, i.e., $p_{c}\sim1/N$\ \citep{Bollobas},
see Fig.\ \ref{fig:diluteSKphase}.)

The following results are obtained with the Extremal Optimization
heuristic (EO)\ \citep{Boettcher00,Boettcher01a,Dagstuhl04}. For
a generic combinatorial optimization problem, EO performs a local
search\ \citep{Hoos04,Dagstuhl04} on an existing configuration of
$N$ variables by changing preferentially those of poor \textit{local}
arrangement. For example, in case of the spin glass model in Eq.\ (\ref{eq:Hamiltonian}),
it assigns to each spin variable a ``fitness'' $\lambda_{i}=\sigma_{i}\sum_{j=1}^{N}J_{i,j}\sigma_{j}$,
corresponding to the negative of the local energy of each spin, so
that $H=-\frac{1}{2}\sum_{i=1}^{N}\lambda_{i}$ reproduces the Hamiltonian
for SK in Eq.\ (\ref{eq:Hamiltonian}). A local search with EO requires
the ranking of these fitnesses $\lambda_{i}$ from worst to best,
$\lambda_{\Pi(1)}\leq\lambda_{\Pi(2)}\leq\ldots\leq\lambda_{\Pi(N)}$,where
$\Pi(k)=i$ is the index for the $k^{{\rm th}}$-ranked variable $\sigma_{i}$.
In the basic version of EO , it always updates the lowest rank, $k=1$\ \citep{Boettcher00,CISE,Brownlee11}.
Instead, $\tau-$EO as used here selects the $k^{{\rm th}}$-ranked
variable with a \emph{scale-free} probability $P_{k}\propto k^{-\tau}$.
The selected variable is updated \emph{unconditionally}, and its fitness
and that of its neighboring variables are reevaluated. This update
is repeated as long as desired, where the unconditional update ensures
significant fluctuations, yet, sufficient incentive to return to near-optimal
solutions due to \emph{selection against} variables with poor fitness,
for the right choice of $\tau$. Clearly, for finite $\tau$, EO never
``freezes'' into a single configuration; it instead records one
(or even an extensive set\ \citep{Boettcher04a,Boettcher03a}) of
the best configurations in passing. Our specific implementation of
$\tau-$EO for SK proceeds is described in Ref.~\citep{EOSK}. 

EO is implemented\ \footnote{See also the demo at http://www.physics.emory.edu/fac-ulty/boettcher/Research/EO\_demo/demoSK.c}
for denser instances ($p\geq0.05$) as described in Refs.\ \citep{EOSK,Boettcher10b},
for sparser instances ($p\leq0.05$) as described in Refs.\ \citep{Boettcher03a,Boettcher03b};
we have obtained statistically identical results for \emph{both} at
$p=0.05$. For any given value of $p$, we generate a large number
of instances over a large range of sizes $N$ (from $10^{5}$ instances
for all $N<200$ to $2\times10^{3}$ at $N\approx1000$, to $10^{2}-10^{3}$
for $N>1000$) and average the obtained ground-state energies, $\left\langle e_{0}\right\rangle _{N}$,
plotted as a function of $N$ in Fig.\ \ref{fig:diluteSKextra}a-l.
It is well-known that finding solutions of lowest energy for each
instance corresponds an NP-hard combinatorial problem (Max-Cut \citep{G+J}),
and a significant effort must be undertaken to minimize systematic
errors in the approximation of ground states. Luckily, we can gauge
the accuracy of EO (and any other heuristic \citep{Boettcher19})
using the theoretical predictions already obtained with RSB. For instance,
in panel (a) of Fig.\ \ref{fig:diluteSKextra}, pertaining to SK
($p=1$) as previously studied in Ref.\ \citep{Boettcher10b}, the
EO data was extrapolated to the thermodynamic limit and fit according
to Eq.\ (\ref{eq:FSC}) to reproduce the RSB prediction for $e_{{\rm Parisi}}$
to 5 digits of accuracy. Similarly, EO applied to sparse networks\ \citep{Boettcher03a,Boettcher03b}
reproduced the RSB prediction for BL of fixed degree $c=3$ from Ref.\ \citep{mezard:01}
to 4 digits of accuracy. Further application of EO to BL of fixed
degrees $c=4,\ldots,26$ provided predictions for thermodynamic $\left\langle e_{0}^{(c)}\right\rangle _{N=\infty}$,
which themselves extrapolate consistently for $c\to\infty$ such that
$c^{-\frac{1}{2}}\left\langle e_{0}^{(c)}\right\rangle _{\infty}\sim e_{{\rm Parisi}}$.
Thus, the extrapolation plot, i.e., the very fact that a scaling according
to Eq.\ (\ref{eq:FSC}) can be consistently applied, becomes a bootstrap
measure of validation in its own right\ \citep{Boettcher19}.

In Tab.\ \ref{datatable} we list all parameters obtained from the
data displayed in Fig.\ \ref{fig:diluteSKextra} for each value of
$p$ asymptotically for large $N$ to Eq.\ (\ref{eq:FSC}). We observe
that the dependence of the FSC exponent $\omega$ on $p$, shown in
Fig.\ \ref{fig:omegaSKplot}, is quite remarkable. While SK \citep{EOSK,Aspelmeier07,Kim07,Boettcher10b}
as well as sparse networks \citep{Boettcher03a,Boettcher03b,Bouchaud03}
with bimodal bonds have consistently exhibited FSC with $\omega\approx\frac{2}{3}$,
independent of degree $c$, for fixed $p<1$ in the dilute SK we find
significant variation in $\omega(p)$. For decreasing $p$, $\omega(p)$
rises from its SK-value at $p=1$ with what appears to be a continuous
hyperbolic form, roughly $\omega-\frac{2}{3}\sim\frac{1}{p}$, for
about two decades, $0.03\leq p\leq1$, as the inset of Fig.\ \ref{fig:omegaSKplot}
suggests. Leaving $\left\langle e_{0}\right\rangle _{\infty}$ as
a fitting paramters, the exact result, $\left\langle e_{0}\right\rangle _{\infty}=e_{{\rm Parisi}}$,
is reproduced within errors for $p>0.03$, see Tab.\ \ref{datatable}.
Significant deviations only arise for the smallest values of $p$
studied here, and it is not obvious whether these are due to systematic
errors in EO or due to the assumptions underlying Eq.\ (\ref{eq:FSC})
Fixing $\left\langle e_{0}\right\rangle _{\infty}=e_{{\rm Parisi}}$
for the fit has virtually no effect on our key result, the variation
of $\omega\left(p\right)$, as listed in Tab.\ \ref{datatable},
for $p>0.03$. However, the data for smaller $p$ no longer fit to
any FSC we considered, such as higher order corrections to Eq.\ (\ref{eq:FSC}),
logarithmic corrections, etc., unless we assume large systematic errors
and discount EO-data for larger $N$. Not only does that contradict
aforementioned results in Refs. \citep{Boettcher03a,Boettcher03b},
it renders any such fit arbitrary. It is interesting that this transition
occurs at a value of $p$ where the fitted value of $\omega(p)$ just
about passes unity, suggesting that a bulk ($1/N$) correction, subdominant
in Eq.\ (\ref{eq:FSC}) for $\omega<1$, might interfere. (At $p=1$,
such a correction has been ruled out for SK in Ref.\ \citep{Boettcher10b}.)
The breakdown of simple FSC in Eq.\ (\ref{eq:FSC}) is also signaled
by the diverging amplitude $A\left(p\right)$ in Tab.\ \ref{datatable}. 

An analytic study of the dilute SK model in the limit of $p\to0$
should be able to reveal whether the limit for $\omega$ is regular
or singular. A perturbative expansion around that limit might also
shed light on the nature of FSC in RSB, since there does not appear
to be a transition at any finite $p$ from RSB near $p=1$ to a simple
replica-symmetric phase, at least at $T=0$. Thus, future studies
should explore the properties of the dilute SK for finite $T$. But
even at the ground-state level, we intend to explore the behavior
of other characteristic features, such as the ensemble fluctuations
in the ground state energies \citep{Parisi08,Parisi09,Boettcher10b}.
As there is expected to be a scaling relation between the FSC-exponent
$\omega$ and the exponent describing such fluctuations \citep{Bouchaud03},
investigating their relation while evolving with $p$ should be very
revealing about the nature of universality at $T=0$.

\bibliographystyle{apsrev}
\bibliography{../../../../../stb/Boettcher}

\end{document}